\documentstyle[sprocl]{article}

\input{psfig}
\def\Journal#1#2#3#4{{#1} {\bf #2}, #3 (#4)}

\def\MN{\em M.N.R.A.S.} 
\def\APJ{\em Astophys. J.}

\def\be{\begin{equation}}
\def\ee{\end{equation}}
\def\bea{\begin{eqnarray}}
\def\eea{\end{eqnarray}}
\begin{document}

\title{THE 2dF GALAXY REDSHIFT SURVEY: Preliminary Results} 

\author{ STEVE MADDOX \footnote[1]{ 
{\sl On behalf of the 2dF Galaxy Redshift Survey Team:} 
Steve Maddox (IoA), Joss Bland-Hawthorn (AAO), Russell Cannon (AAO),
Shaun Cole (Durham), Matthew Colless (MSSSO), Chris Collins (LJMU),
Warrick Couch (UNSW), Gavin Dalton (Oxford), Simon Driver (UNSW),
Richard Ellis (IoA), George Efstathiou (IoA), Simon Folkes (IoA),
Carlos Frenk (Durham), Karl Glazebrook (AAO), Nick Kaiser (AAO/CITA),
Ofer Lahav (IoA), Stuart Lumsden (AAO), Bruce Peterson (MSSSO), John
Peacock (ROE), Will Sutherland (Oxford), Keith Taylor (AAO).}
}
\address{ Institute of Astronomy, Cambridge, UK}


\maketitle\abstracts{
Spectroscopic observations for a new survey of 250 000 galaxy redshifts
are underway, using the 2dF instrument at the AAT.  The input galaxy
catalogue and commissioning data are described. The first result
from the preliminary data is a new estimate of the galaxy luminosity
function at $\langle z \rangle = 0.1$.  }
  
\section{Introduction} 

At the end of October 1997, the AAO 2dF instrument (Cannon, these
Proceedings) is working reliably with the full 400 fibre system,
and the commissioning phase is coming to an end.
We are now starting the spectroscopic observations for a redshift survey 
of 250 000 galaxies, covering 2000 $\Box^2$ 
in the southern hemisphere. 
The survey will have a mean depth of $z=0.1$, and a volume of $1-2
\times 10^7 h^{-3} {\rm Mpc }^3$, which is  $\sim 10$ times larger than any
present redshift survey (see Fig.~1). %\ref{fig:sky}(a)).
The large volume and dense spatial sampling will enable us to quantify
statistics of the galaxy distribution on large scales. 
Updated information about the survey can be found at
 {\tt http://www.ast.cam.ac.uk/\~\,sjm/ }

\begin{figure}
\hspace*{1cm} 
\psfig{figure=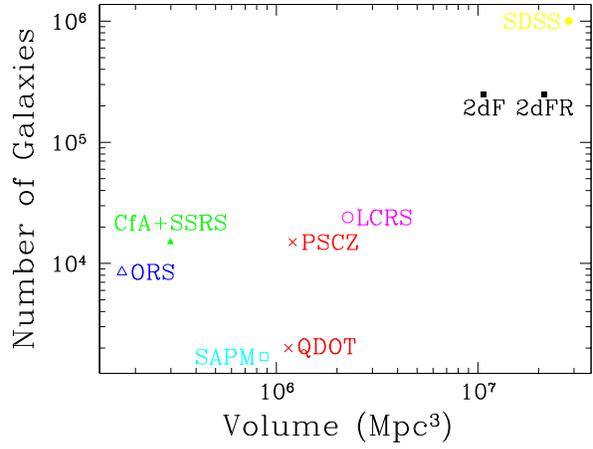,height=7cm,angle=-90} 
\caption{Comparison of volume and number of galaxies to other
surveys. }
\label{fig:surveys}
\end{figure}
\begin{figure}
\psfig{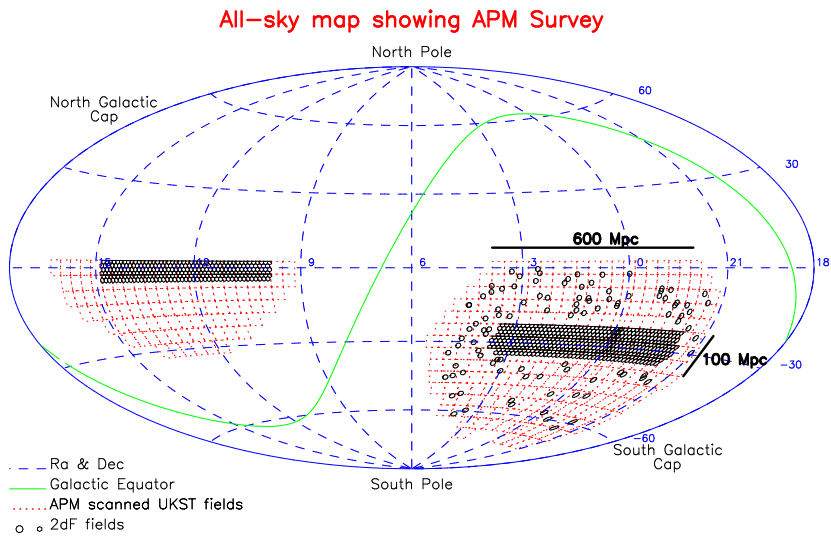}
\caption{An equal area plot of the available APM data, and proposed 
2dF survey area. }
\label{fig:sky}
\end{figure}

\section{The Photometric Catalogue}   

The galaxies are selected from APM scans of UKST Bj plates,
analyzed using techniques described by Maddox et
al~\cite{me90b}$^{\!,\,}$\cite{me96}.
Accurate positions of galaxies and fiducial stars are vital to avoid
light loss outside the fibres, so we have improved the astrometric
calibration by correcting for systematic distortions in
the Schmidt plates.
In addition, APM scans of UKST Second Epoch R plates are used to
select fiducial guide stars which have negligible proper motion.  These
R-plates also provide broad-band colours for the galaxies.

We chose to observe 2000 $\Box^2$ of the revised APM catalogue
at high galactic latitude, where there is little obscuration from
Galactic dust. We estimated dust extinction from recalibrated,
temperature corrected IRAS 100 $\mu m$ emission maps~\cite{schlegel}.
The survey area consists of: a contiguous block $15^\circ \times
75^\circ$ in the SGP; a contiguous block $7.5^\circ
\times 75^\circ$ in the NGP; and 100 randomly positioned 2dFs spread
over the SGP (see Fig.~2). %\ref{fig:sky}(b)).
The magnitude limit is $b_J \le 19.5,$ set to match the surface density
of galaxies to the 400 fibres per 2dF. At this limit, the integration
time required to obtain a redshift ($\sim 45 {\rm mns}$) is roughly
matched to the expected time to reconfigure the fibres for the next
field.

Allowing for the average weather statistics, the galaxy survey
will require 80 nights to complete. We merged our target list with
the UVX QSO sample of Boyle et al~\cite{boyle}, which was selected
from other APM data in the same astrometric system. 
 The combined survey requires 90 nights.  In
addition, we aim to use the best 10\% of observing conditions to
measure redshifts for a sample of $6000$ galaxies $R\le 21$.  These
galaxies are selected from APM scans of UKST 4415 films and should
have $\langle z \rangle \sim 0.3 $.  Our first scheduled
PATT time is planned for January 1998, and we hope to
finish the surveys by the end of 2000.

\begin{figure}
\begin{minipage}{5.5cm}
\psfig{figure=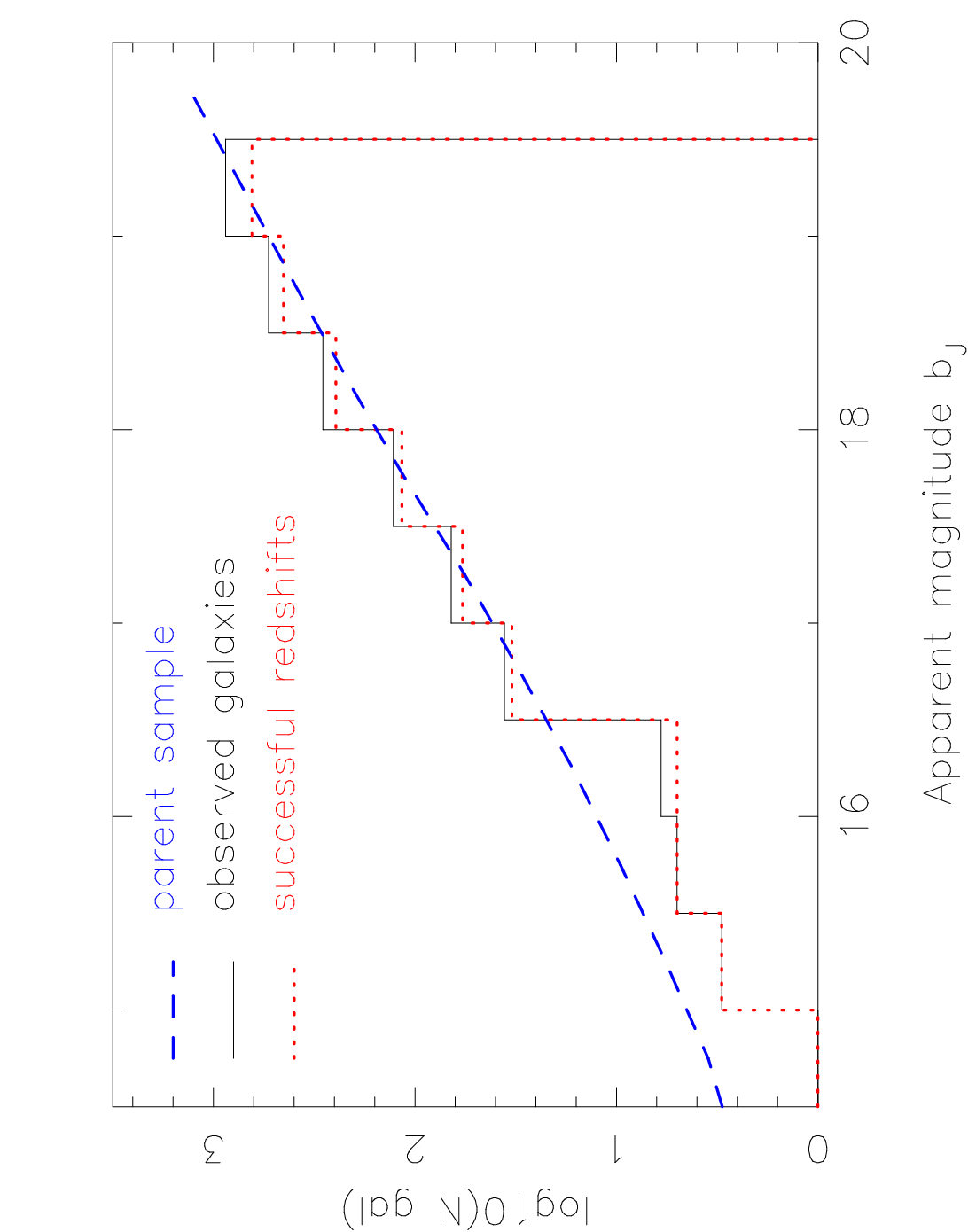,height=7.5cm,angle=0}
\caption{$N(b_J)$ of the parent sample (\hbox{- - -}), observed galaxies 
(---) and successful redshifts ($^{.....}$). }
\end{minipage}
\label{fig:nm}
\begin{minipage}{5.5cm}
\vspace*{-0.3cm}
\psfig{figure=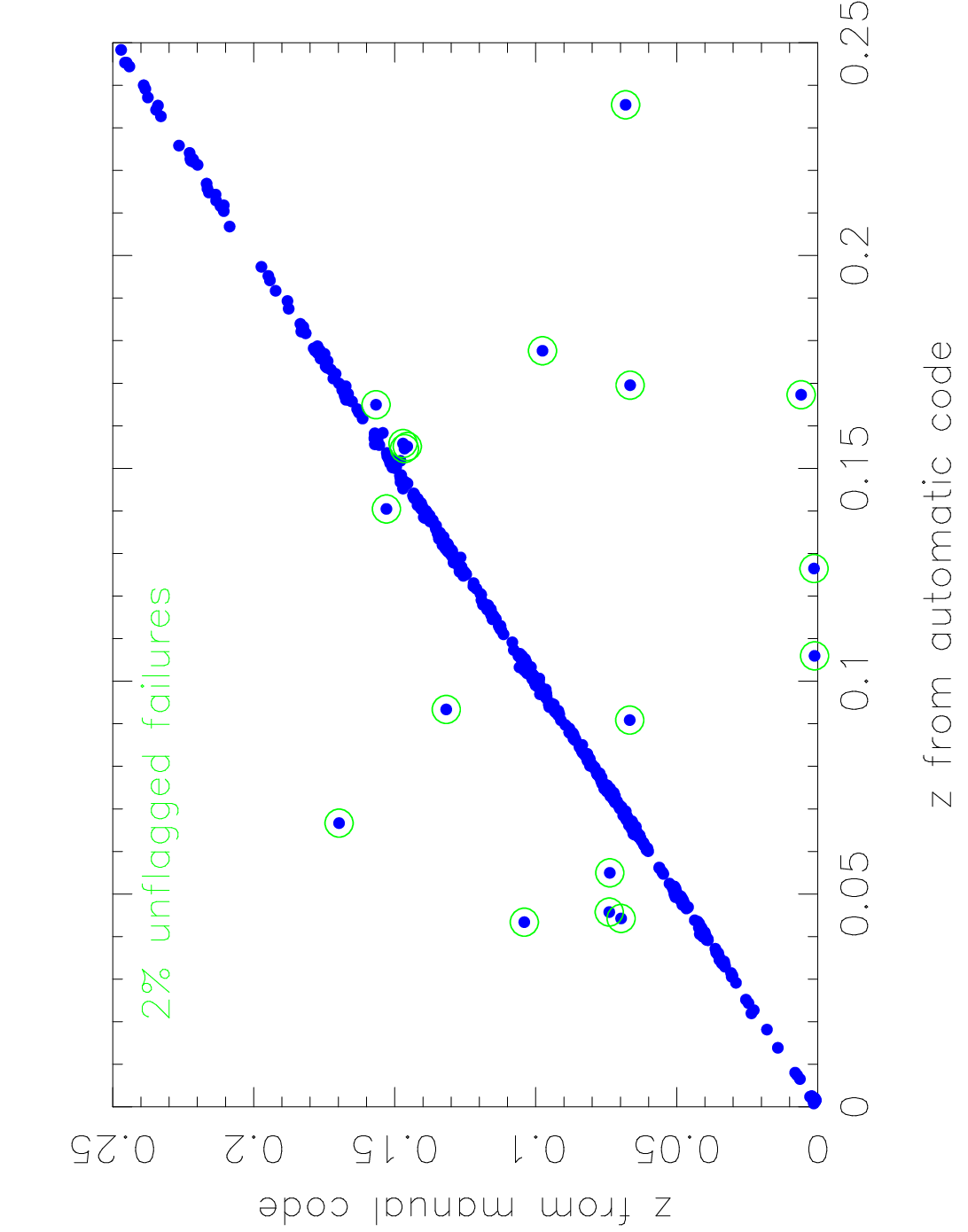,height=7.5cm,angle=0}
\caption{Comparison between manual and automatic redshift 
measurements. }
\end{minipage}
\label{fig:z1z2}
\begin{minipage}{5.5cm}
\psfig{figure=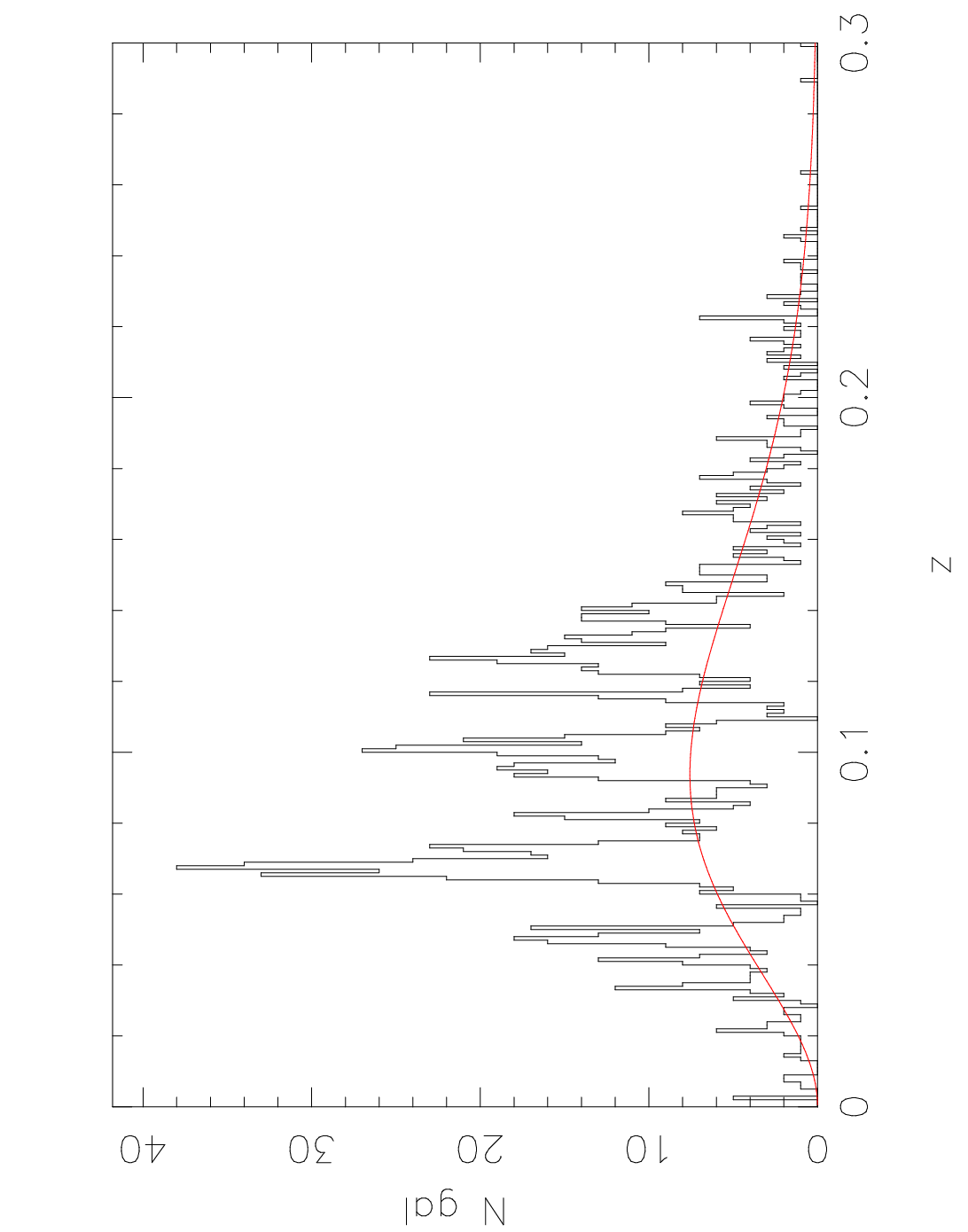,height=7.5cm,angle=0}
\caption{$N(z)$ for 2dF galaxies. The line
shows the distribution from Maddox et al$^2$}
\label{fig:nz}
\end{minipage}
\hspace*{0.5cm}
\begin{minipage}{5.cm}
\psfig{figure=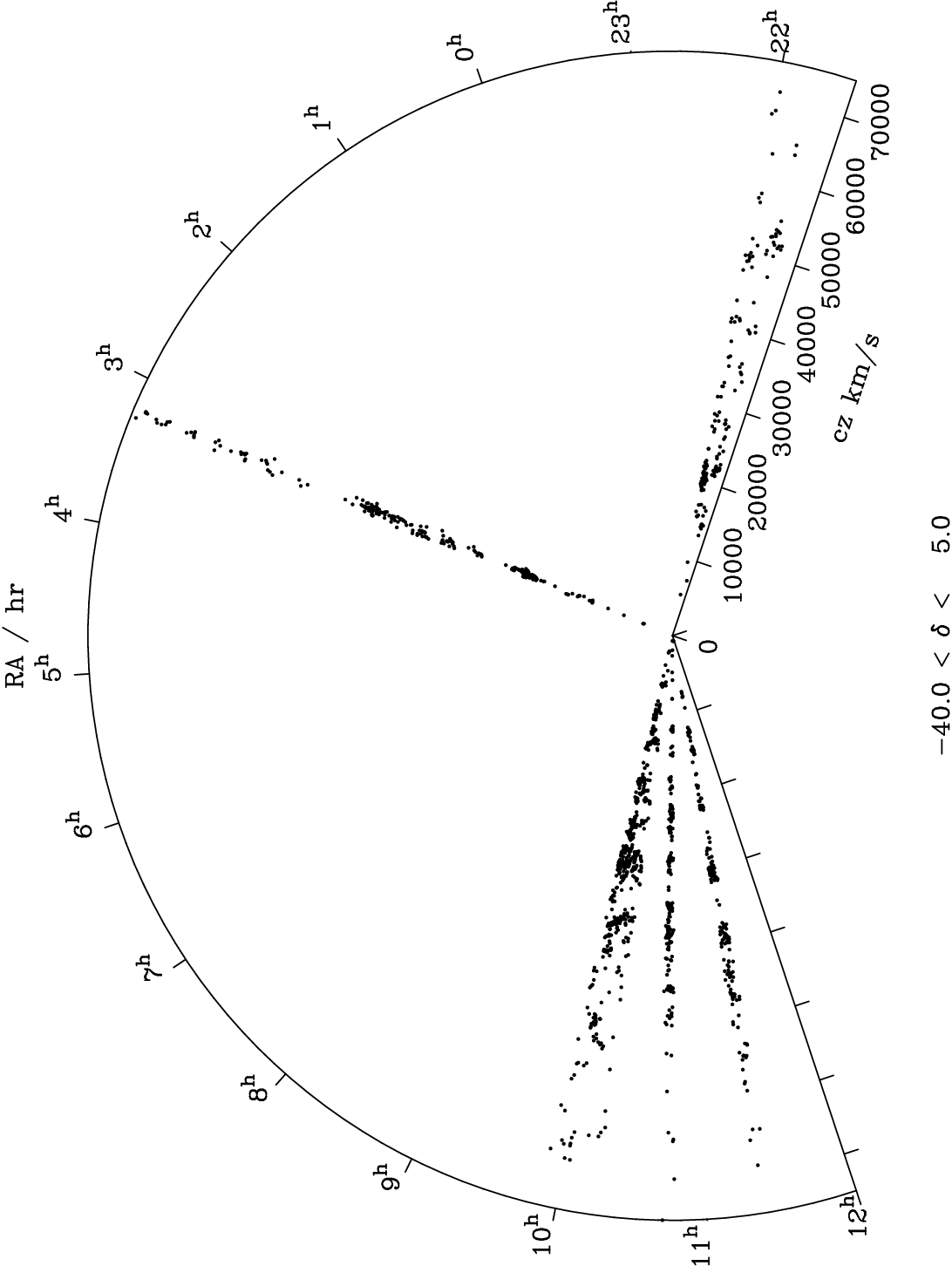,height=7cm,angle=0}
\caption{Cone-plot of 2dF galaxies.}
\end{minipage}
\label{fig:cone}
\end{figure}

\section{Scientific Goals}

We will use the survey to address a variety of fundamental problems 
in galaxy formation and cosmology. Some specific measurements are: 

\noindent$\bullet$ The power spectrum of galaxy
clustering on scales $\sim 100 h^{-1}$ Mpc. This allows a direct
comparison between galaxy fluctuations and microwave background
anisotropies on the same spatial scales.

\noindent$\bullet$ The distortion of the clustering pattern in
redshift space. This measures $\Omega^{0.6}/b$ and so constrains the 
cosmological density parameter $\Omega$ and the spatial distribution
of dark matter.

\noindent$\bullet$ The morphology of galaxy clustering. The topology
and shape of the density distribution can test whether the initial
fluctuations are Gaussian, as predicted by inflationary models of the
early universe.

\noindent$\bullet$ Correlations between galaxy properties and
large-scale structures, and  variations in the
clustering and velocity distributions of galaxies as a function of
their luminosity, type and star-formation history. These two
complementary approaches quantify environmental effects which will
provide crucial constraints on models of galaxy formation.

\noindent$\bullet$ The objective classification of spectra using
Principal Components Analysis and Artificial Neural Nets. These will 
give physical insight into the Hubble sequence. 

\noindent$\bullet$  The evolution of the galaxy luminosity
function, star-formation rate and galaxy clustering amplitude out to a
redshift of $z \sim 0.5$. 

\noindent$\bullet$  Clusters and groups of galaxies, and in particular
of the infall in clusters and dynamical estimates of cluster masses at
large radii.

\begin{figure}
\hspace*{2cm} 
\psfig{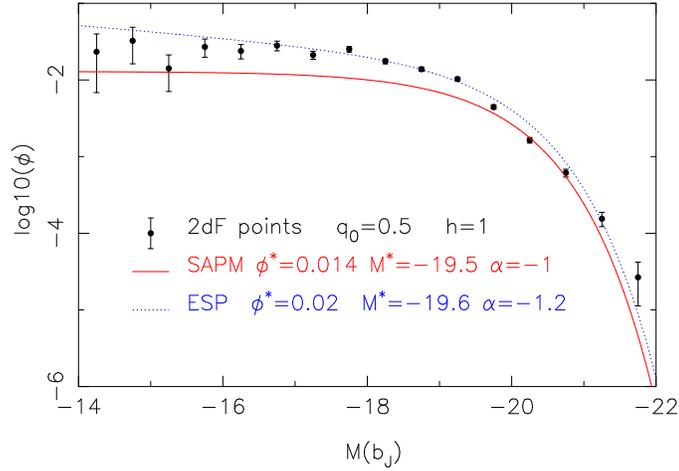}
\caption{Preliminary luminosity function of 2dF galaxies compared to
the Stromlo/APM, and the ESO Slice Project estimates. The points show
the $V/V_{max}$ estimate, and the errors show the Poisson
counting error on the number of galaxies in each magnitude bin.}
\label{fig:phi}
\end{figure}

\section{ First Results} 

Up to the end of June 1997, 16 fields of 200 fibres were observed for
the galaxy survey during commissioning.  1937 spectra were reduced
automatically using the AAO 2dFdr pipe-line software and some also
interactively using IRAF.
We obtained 782 redshifts by hand, and 1445 by 
automatic cross-correlation, emission line detection and pca
analysis. This gave a total of 1562 galaxies with reliable redshifts. 

The commissioning data are (not surprisingly) of variable quality,
and the average completeness is only 80\%. There is very little dependence
on magnitude (see Fig.~3). %\ref{reds}(a)). 
On fields which are as good 
as we expect for the survey, the completeness is $\sim 90\%$.
The unsupervised automatic codes are already highly reliable, giving
only  $\sim $2\%  undetected failures (see Fig.~4), and we expect this
to improve as instrument and software are tuned. 
%\ref{reds}). 
The rms velocity error estimated by comparing multiple measurements
for 646 galaxies is $\simeq 140 \ {\rm kms^{-1}} $. 

The mean redshift is $\langle z \rangle = 0.1$ consistent with our
expected value (see Fig.~5). %\ref{reds}).  
In Fig.~6 %\ref{reds} 
the galaxy positions in a redshift space cone plot show strong
clustering, with some large structures apparently extending between widely
separated 2dFs.  

We calculated a preliminary luminosity function, using the standard
$V/V_{max}$ estimate, and the ratio of the observed to parent
magnitude distribution to correct for incompleteness. As seen in
Fig~7, %\ref{phi}, 
the normalization is consistent with that found by
the ESO slice project~\cite{esp}. This is higher than the
SAPM~\cite{sapm} result at $\langle z \rangle = 0.05$, even though the
SAPM and 2dF samples are selected from essentially the same galaxy
catalogue. This strongly argues against the SAPM measurement being low
because of simple magnitude errors in the APM catalogue, as has been
suggested by some authors. A combination of a local underdensity and
evolution is the most probable explanation, though surface-brightness
selection effects may also be important. 

\section*{ The Latest Update} 
At the beginning of November 1997, 4 nights of observing with variable
weather yielded over 3000 spectra, bringing the current total up to
5000. For the first time we achieved more than 1000 redshifts in one
night of observing. The data are already reduced and the resulting
luminosity function agrees well with the result from the commissioning
data.


\section*{References}

\begin{thebibliography}{99}
\bibitem{me90b} Maddox, S.J., Sutherland, W.J., Efstathiou, G., and
Loveday J.
 \Journal{\MN}{243}{692}{1990}
\bibitem{me96} Maddox, S.J., Efstathiou G.P. and Sutherland
 W.J. \Journal{\MN}{283}{1227}{1996} 
\bibitem{schlegel} Schlegel, Finkbeiner, D.,  Davis, M., astro-ph/9710327 
\bibitem{boyle} Boyle, B., Smith, R.J., Shanks, T., Croom, S.M., 
Miller, L., astro-ph9710202
\bibitem{sapm} Loveday, J., Peterson, B.A., Efstathiou, G. and 
Maddox, S.J. \Journal{\APJ}{390}{338}{1992}
\bibitem{esp} E. Zucca, G. Zamorani, G. Vettolani, A. Cappi,
R. Merighi, M. Mignoli, G. M. Stirpe, H. MacGillivray, C. Collins,
C. Balkowski, V. Cayatte, S. Maurogordato, D. Proust, G. Chincarini,
L. Guzzo, D. Maccagni, R. Scaramella, A. Blanchard, M. Ramella 
\Journal{A\&A}{390}{338}{1997}

\end{thebibliography}
\end{document}